\documentclass[twocolumn]{aastex631}

\newcommand{\Eq}[1]{Eq.~\ref{#1}}
\newcommand{\Fig}[1]{Fig.~\ref{#1}}
\newcommand{\Sec}[1]{Section~\ref{#1}}
\newcommand{\App}[1]{Appendix~\ref{#1}}
\newcommand{\Tab}[1]{Table~\ref{#1}}
\newcommand{\threehun}{{\sc The Three Hundred}}
\newcommand{\tth}{{\sc The300}}
\newcommand{\omsig}{$\Omega_{\rm{M}}$ and $\sigma_{8}$}

\DeclareRobustCommand{\VAN}[3]{#2}
\let\VANthebibliography\thebibliography
\def\thebibliography{\DeclareRobustCommand{\VAN}[3]{##3}\VANthebibliography}

\usepackage{multirow}
\usepackage{booktabs}

\begin{document}

\title{Constraining cosmological parameters using the splashback radius of galaxy clusters}

\newcommand{\affuwaterloo}{Department of Physics and Astronomy, University of Waterloo, Waterloo, Ontario N2L 3G1, Canada}
\newcommand{\affwca}{Waterloo Centre for Astrophysics, University of Waterloo, Waterloo, Ontario N2L 3G1, Canada}
\newcommand{\affuedin}{Institute for Astronomy, University of Edinburgh, Royal Observatory, Edinburgh EH9 3HJ, UK}
\newcommand{\afficrar}{International Centre for Radio Astronomy Research, University of Western Australia, 35 Stirling Highway, \\Crawley, Western
Australia 6009, Australia}
\newcommand{\affastrod}{ARC Centre of Excellence for All-Sky Astrophysics in 3 Dimensions (ASTRO 3D), Australia}

\author[0000-0001-5490-9621]{Roan Haggar}
\altaffiliation{Corresponding author. Email: rhaggar@uwaterloo.ca}
\affiliation{\affuwaterloo}
\affiliation{\affwca}

\author{Yuba Amoura}
\affiliation{\affuwaterloo}
\affiliation{\affwca}

\author{Charlie T. Mpetha}
\affiliation{\affuwaterloo}
\affiliation{\affwca}
\affiliation{\affuedin}

\author{James E. Taylor}
\affiliation{\affuwaterloo}
\affiliation{\affwca}

\author{Kris Walker}
\affiliation{\afficrar}
\affiliation{\affastrod}

\author{Chris Power}
\affiliation{\afficrar}
\affiliation{\affastrod}



\begin{abstract}

Cosmological parameters such as \omsig\ can be measured indirectly using various methods, including galaxy cluster abundance and cosmic shear. These measurements constrain the composite parameter $S_{8}$, leading to degeneracy between \omsig. However, some structural properties of galaxy clusters also correlate with cosmological parameters, due to their dependence on a cluster's accretion history. In this work, we focus on the splashback radius, an observable cluster feature that represents a boundary between a cluster and the surrounding Universe. Using a suite of cosmological simulations with a range of values for \omsig, we show that the position of the splashback radius around cluster-mass halos is greater in cosmologies with smaller values of $\Omega_{\rm{M}}$ or larger values of $\sigma_{8}$. This variation breaks the degeneracy between \omsig\ that comes from measurements of the $S_{8}$ parameter. We also show that this variation is, in principle, measurable in observations. As the splashback radius can be determined from the same weak lensing analysis already used to estimate $S_{8}$, this new approach can tighten low-redshift constraints on cosmological parameters, either using existing data, or using upcoming data such as that from {\textit{Euclid}} and LSST.

\end{abstract}



\section{Introduction}
\label{sec:intro}

Over the past few decades, multiple probes have converged on a single ``concordant'' cosmological model, the 6-parameter $\Lambda$-Cold Dark Matter (LCDM) model. Despite its successes, however, the origin and fundamental nature of the main ingredients of this model -- inflation, dark energy, and dark matter -- remain unexplained. Thus, there is a current emphasis on exploring tensions between independent cosmological tests, that might hint at fundamental revisions to the model. 

In structure formation, there is a notable tension between independent determinations of the $\sigma_8$ parameter, the amplitude of the linear power spectrum on the scale of $8\,h^{-1}\,$Mpc, based on analyses of Cosmic Microwave Background (CMB) fluctuations \citep{Planck2020} and cosmic structure at low redshift. Specific examples of low-redshift measurements include those from weak gravitational lensing \citep[e.g.][]{Abbott2022, Li2023,Dalal2023,Burger2024}, galaxy cluster abundance \citep[e.g.][]{Lesci2022,Ghirardini2024,Aymerich2024,Bocquet2024}, redshift space distortions \citep[e.g.][]{Nunes2021,Benisty2021}, and galaxy clustering or CMB lensing tomography \citep[e.g.][]{Krolweski2021,White2022,Philcox2022,Marques2024}. We refer the reader to \citet{DiValentino2021,Perivolaropoulos2022,Abdalla2022} for reviews.

As highlighted in \cite{Preston2023}, these measurements differ both in redshift and in spatial scale; scale-dependent modifications to the LCDM model can reconcile low-redshift measurements to CMB constraints. Given CMB-based measurements have already reached their limiting precision, further tests of scale or redshift dependence require improved constraints at low redshift.

Low-redshift tests of structure formation generally constrain the present-day amplitude of fluctuations corrected for the effect of dark energy, parameterized by the $S_{8}$ parameter,
\begin{equation}
     S_{8}=\sigma_{8}\left(\frac{\Omega_{\rm{M}}}{0.3}\right)^{0.5}\,.
     \label{eq:s8}
\end{equation}
As $S_{8}$ is dependent on both $\sigma_{8}$ and the matter density parameter, $\Omega_{\rm{M}}$, measurements of this parameter introduce a degeneracy between higher-density, smoother models (greater $\Omega_{\rm{M}}$, smaller $\sigma_{8}$) and lower-density models with larger fluctuations. However, other properties of cosmological structure vary differently with \omsig. In particular, over a broad range of masses and redshifts, the half-mass assembly redshift \citep{amoura2021} and instantaneous growth rate \citep{amoura2024} of galaxy groups and clusters vary almost orthogonally in the $\Omega_{\rm{M}}$-$\sigma_8$ plane to measurements of $S_{8}$ from weak lensing and cluster abundance. While quantities such as the age and growth rates of clusters are not directly observable, they leave signatures that can serve as observable proxies or estimators, such as the internal structural properties of clusters \cite[e.g. Haggar et al.~in prep;][]{wu2013,gouin2021,amoura2024}. 

The outskirts of clusters, the so-called ``infall regions'', are particularly strongly impacted by the recent accretion of material from the surrounding density field. This region consists of material being accreted into the cluster for the first time, and also material that has turned around, and is on a second infall towards the cluster center \citep{haggar2023}. This is indicated by the presence of a ``splashback radius'' \citep{diemer_sparta2017}, which represents the point at which recently-accreted bound material reaches its first apocenter, such that the density beyond this point is instead dominated by material currently approaching the cluster for the first time. The splashback feature has been studied in simulations \citep{adhikari2014,diemer2014} and observations \citep{chang2018,zurcher2019,rana2023}, both of which show that it corresponds to the point at which the density profile drops most steeply. Further studies of the splashback radius also indicate that this feature depends on the relative accretion rate of clusters \cite[e.g.][]{Diemer2017,Fong2021,shin2023}. Consequently, as the growth rate of structure depends on cosmological parameters like \omsig, the splashback radius should also exhibit a cosmological dependence.

In this paper, we explore the cosmological information contained in measurements of the infall region, and the splashback radius in particular. We do this by studying the splashback feature in a suite of simulations with different values of \omsig, and showing how this observable quantity correlates with cosmological parameters that are not directly observable. The outline of the paper is as follows: in \Sec{sec:methods} we describe the simulation data and methods, in \Sec{sec:results} we present our results, and in \Sec{sec:discussion} we discuss the physical interpretation of these findings. Finally, in \Sec{sec:conclusions}, we summarize our results and discuss how they might be used to observationally constrain these cosmological parameters.

\section{Simulations and methodology}
\label{sec:methods}

To study the variation of cluster structure with cosmology, we have run a set of 21 $1024^3$-particle dark matter-only N-body simulations for a range of values of \omsig. A lower-resolution subset of these simulations are described in \citet{amoura2021} and \citet{amoura2024}, and the full suite of simulations will be described extensively in an upcoming paper (Amoura et al.~in prep), but we summarize their main properties here.

\subsection{Simulation data}
\label{sec:simulations}

All simulations were run the N-body code GADGET-4\footnote{\url{https://wwwmpa.mpa-garching.mpg.de/gadget4/}} 
\citep{springel2021}, with a box size of $500\,h^{-1}\,$Mpc (comoving), where $h$ is defined such that the Hubble constant \mbox{$H_{0}=100\,h\,$km$\,$s$^{-1}\,$Mpc$^{-1}$}. The dark matter particle mass for the simulations is equal to \mbox{$m_{\rm{p}} = \Omega_{\rm{M}} \times 3.23 \times 10^{10}\,h^{-1}\,M_\odot$}; for the values of $\Omega_{\rm{M}}$ used, this is equivalent to a typical dark matter particle mass of $\sim10^{10}\,h^{-1}\,M_{\odot}$. This particle mass, along with the softening length \mbox{($\ell_{\rm{s}} = 2.5\,$kpc)} were chosen to obtain a large sample of cluster- and group-sized haloes with sufficient resolution to accurately measure their internal structure. The Amiga Halo Finder \citep[AHF\footnote{\url{http://popia.ft.uam.es/AHF}};][]{gill2004_ahf, knollmann2009} was used to obtain halo catalogues, as well as merger trees for the groups and clusters. 

These simulations use values of $\Omega_{\rm{M}}$ ranging from $0.2$ to $0.4$, and $\sigma_{8}$ from $0.7$ to $1.0$. The exact combinations of these two parameters used in this work are shown in \Tab{tab:simulation_pars}. Each of these simulations use a Hubble parameter of $H_0 = 70\,$km$\,$s$^{-1}\,$Mpc$^{-1}$, baryon density fraction $\Omega_{\rm b}=0.0482$, and spectral tilt $n_{\rm{s}}=0.965$. Our analysis uses halos in the final simulation snapshot, at $z=0$. The initial conditions of each simulation are the same, such that differences in halo profiles between simulations are due to the values of \omsig, not cosmic variance.

\subsection{Splashback radius}
\label{sec:splashback}

\begin{deluxetable}{ccccccc}
	\label{tab:simulation_pars}
	\tablecaption{Combinations of \omsig\ in the 21 simulations used in this work. Cells marked with $``\times"$ indicate that a simulation was run using the corresponding values of \omsig.}
	\tablehead{\multicolumn{2}{r}{ } & \multicolumn{5}{c}{$\Omega_{\rm{M}}$}}
        \startdata
		  \multicolumn{2}{r}{ \ \ \ \ \ \ \ \ \ \ \ \ \ \ } & \ 0.20 \ & \ 0.25 \ & \ 0.30 \ & \ 0.35 \ & \ 0.40 \ \\
		\cmidrule(r){3-7}
            \multirow{7}{*}{$\sigma_{8}$} & 1.00 & $\times$ & & & & \\
            & 0.95 & & & & & \\
            & 0.90 & $\times$ & $\times$ & $\times$ & $\times$ & $\times$ \\
            & 0.85 & & $\times$ & $\times$ & $\times$ & \\
            & 0.80 & $\times$ & $\times$ & $\times$ & $\times$ & $\times$ \\
            & 0.75 & & $\times$ & $\times$ & $\times$ & \\
            & 0.70 & $\times$ & & $\times$ & $\times$ & $\times$ \\
        \enddata
\end{deluxetable}

The splashback radius of a cluster is calculated as the minimum (i.e. most negative) logarithmic slope in the density profile of a cluster. This can be physically interpreted by considering a cluster as a dark matter halo superimposed on a background density field. The slope of an NFW density profile becomes increasingly steep with distance from the halo center, but eventually will plateau at the background density of the surrounding Universe. Consequently, there is some maximally negative slope of the profile.

For each cosmology, we calculate the mean density profile for cluster-sized dark matter haloes, with $M_{200}$ masses greater than $10^{14}\,h^{-1}\,M_{\odot}$. $M_{200}$ is the mass contained within a sphere of radius $R_{200\rm{c}}$ (hereafter $R_{200}$), where the average density inside this sphere is 200 times the critical density of the Universe. We do this by stacking the profiles of these clusters; we first calculate the density of each individual cluster in 400 concentric spherical shells, equally spaced on a logarithmic scale between $0.01\,R_{200}$ and $20\,R_{200}$. The mean density of all clusters is then calculated for each of these shells, providing an average density profile with cluster-centric distances normalized by the $R_{200}$ of each cluster.
 
\begin{figure*}
\includegraphics[width=\textwidth]{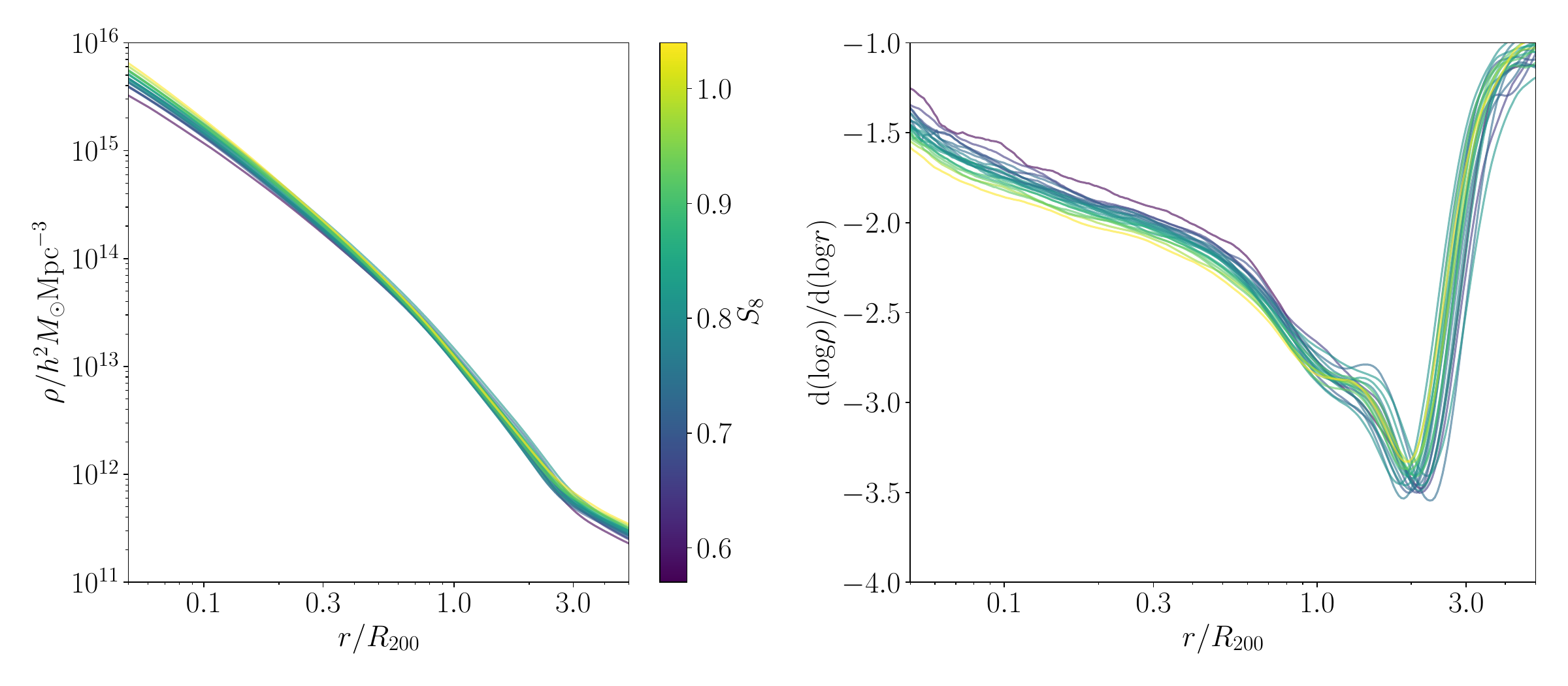}
\caption{Left panel shows the average radial density profiles for haloes of mass $M_{200}>10^{14}\,h^{-1}\,M_{\odot}$ in each of the 21 cosmological simulations described in \Sec{sec:simulations}, where each line corresponds to one simulation. By eye, it can be seen that the greatest variation between cosmologies is in the cluster cores ($r\lesssim0.1R_{200}$) and the splashback region ($r\sim2R_{200}$). Right panel shows the logarithmic slope of these profiles in which a minimum is clearly visible, corresponding to the splashback feature. Color indicates the value of $S_{8}$ in each simulation, defined by \Eq{eq:s8}. Greater values of $S_{8}$ correspond to more collapsed structure and thus more centrally concentrated cluster haloes, and also to a greater background density far from a cluster's influence ($r\gtrsim3R_{200}$). However, there is little correlation between $S_{8}$ and the splashback region in the cluster outskirts.}
\label{fig:density_profiles}
\end{figure*}

We calculate the uncertainty in the mean profile using bootstrapping. For each of the 400 radial regions, we take a sample (with replacement) of size $N$ from the $N$ density values at that radius. We then find the mean density of that sample, and repeat this process 1000 times. The average and uncertainty in the mean density at this radius is taken as the median and $1\sigma$ spread in these 1000 values. This process is repeated independently for all 400 radial bins. The slope in log space of this average profile, ${\rm{d(log}}\rho)/{\rm{d(log}}r)$ is then calculated numerically. The 21 density profiles (for each of the 21 combinations of \omsig), and their slopes, are shown in \Fig{fig:density_profiles}.

For each cosmology, there is a clearly visible minimum at a distance of $\sim2R_{200}$, but the exact location of this minimum varies between the simulations. The position of each of these minima and the uncertainty in their positions were calculated by fitting a quadratic curve to the data around each minimum. In \Fig{fig:density_profiles} the density profiles are smoothed for the sake of readability, but an example of the unsmoothed profile for one cosmological simulation is given in \App{appendix:splashback_fitting}. It is these raw, unsmoothed profiles that were used in calculating the average cluster splashback radius in each simulation. The reduced-$\chi^{2}$ of each of the quadratic fits is consistent with one, which supports the use of bootstrapping as described above.

\section{Results}
\label{sec:results}

As a function of the cluster radius, $R_{200}$, the average splashback radius for cluster-sized haloes varies by approximately $25\%$ across our suite of simulations. Overall, the ratio between splashback radius, $r_{\rm{sp}}$, and cluster radius, $R_{200}$, increases with $\sigma_{8}$ and decreases with 
$\Omega_{\rm{M}}$. This ratio is maximized in the simulation with $\Omega_{\rm{M}}=0.2$ and $\sigma_{8}=1.0$, where \mbox{$r_{\rm{sp}}=2.26\pm0.01R_{200}$}. Conversely, the ratio between $r_{\rm{sp}}$ and $R_{200}$ is minimized when \mbox{$\Omega_{\rm{M}}=0.4$} and \mbox{$\sigma_{8}=0.7$}, leading to \mbox{$r_{\rm{sp}}=1.83\pm0.02R_{200}$}. The uncertainty in the average ratio between $r_{\rm{sp}}$ and $R_{200}$ is low for all of the combinations of \omsig\ used in this work. This uncertainty varies between $0.01R_{200}$ and $0.04R_{200}$, due to the varying quality of the fit to the density profiles between the simulations.

\begin{figure}
\centering
\includegraphics[width=\columnwidth]{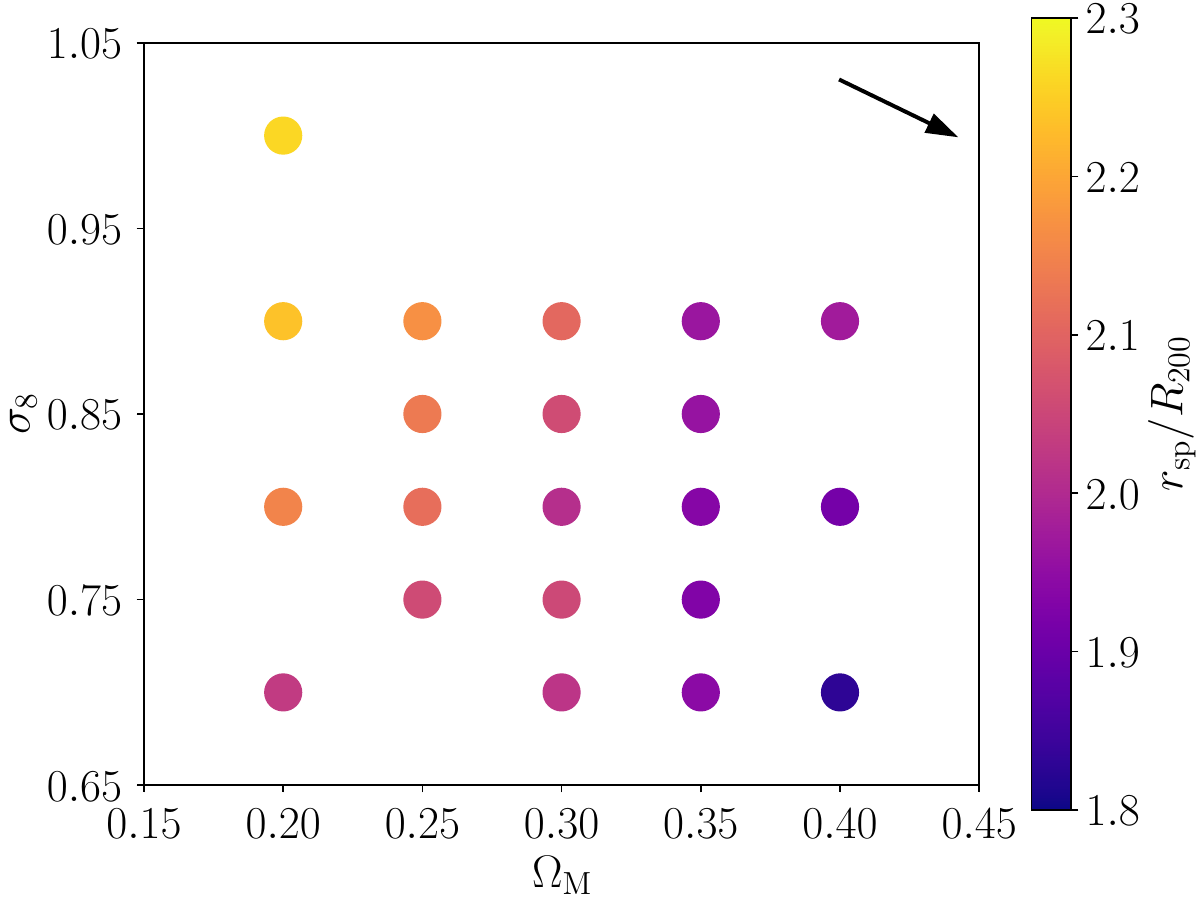}
\caption{Mean splashback radius of cluster-sized haloes as a function of cosmological parameters, for 21 dark matter-only simulations. Color represents the ratio between $r_{\rm{sp}}$ and $R_{200}$, where lighter colors represent a greater splashback radius. Arrow in the top-right corner shows average direction in which $r_{\rm{sp}}$ varies most strongly across the $\Omega_{\rm{M}}$-$\sigma_{8}$ plane.}
\label{fig:splashback_scatter}
\end{figure}

\begin{figure}
\centering
\includegraphics[width=\columnwidth]{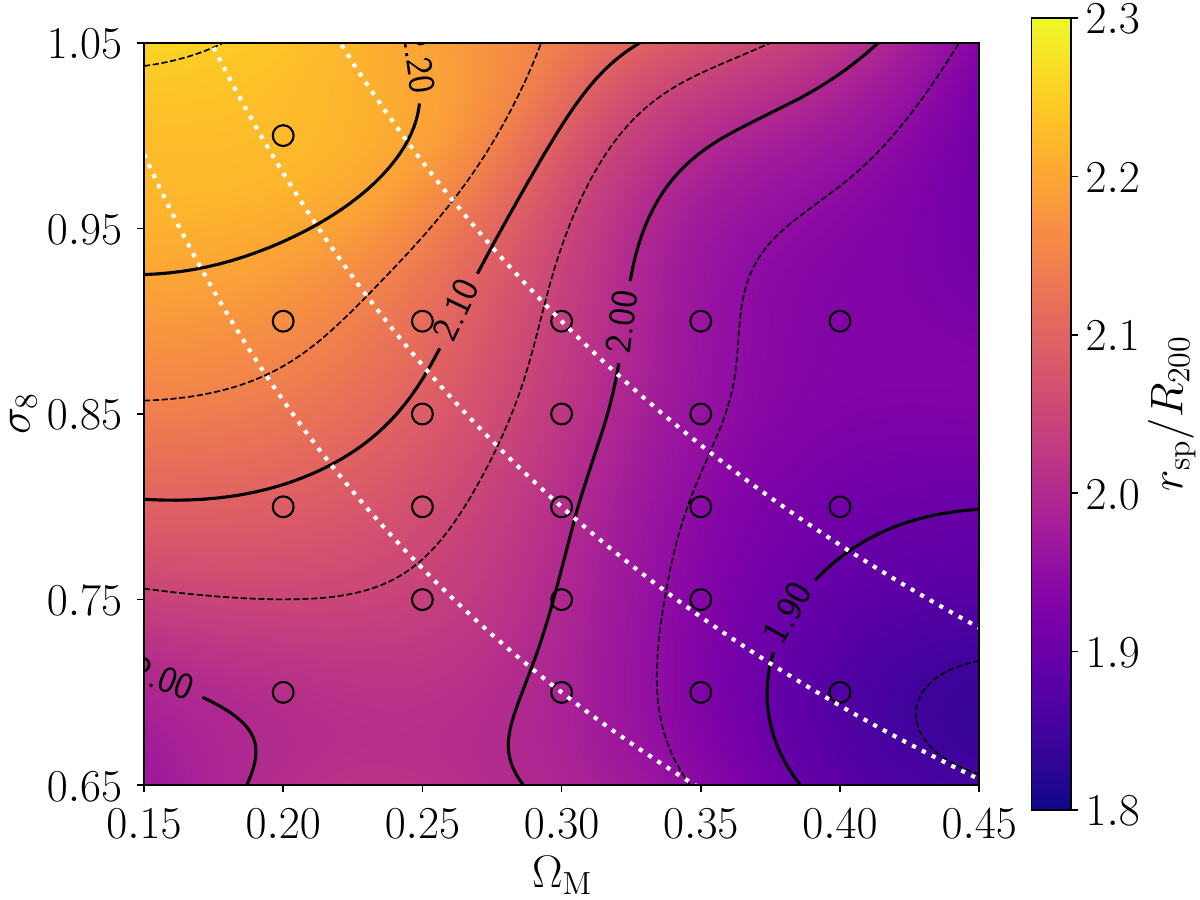}
\caption{Same as \Fig{fig:splashback_scatter}, but using a KDE to interpolate between simulations. Color show the estimated average splashback radius for a galaxy cluster in a Universe with this cosmology. Contours of equal $r_{\rm{sp}}$ are included. Empty circles show the values of \omsig\ for which simulation data exists. White dotted lines show three contours of constant $S_{8}$ for reference, with values of $S_{8}=[0.7, 0.8, 0.9]$.}
\label{fig:splashback_kde}
\end{figure}

These results are summarized in \Fig{fig:splashback_scatter}. Here, each point represents one simulation, and the color shows the mean location of the splashback radius. We also calculate the average direction in this space in which the splashback radius varies most quickly, indicated by the direction of the arrow in the top-right corner. We do this using partial correlation coefficients (PCCs), as described in \citet{lawrance1976}. The partial correlation coefficient between two quantities, $A$ and $B$, whilst controlling for a third quantity, $C$, is equal to
\begin{equation}
    \rho_{AB|C}=\frac{\rho_{AB}-\rho_{AC}\rho_{BC}}{\sqrt{1-\rho_{AC}^{2}}\sqrt{1-\rho_{BC}^{2}}}\,,
    \label{eq:pcc}
\end{equation}
where $\rho_{AB|C}$ is the partial correlation coefficient, and $\rho_{XY}$ is the Spearman's rank correlation coefficient between any two quantities, $X$ and $Y$. Using this, we can independently calculate the correlation coefficients between the splashback radius and $\Omega_{\rm{M}}$, and between the splashback radius and $\sigma_{8}$. The ratio of these two coefficients then provides the average direction in which the splashback radius varies most strongly \citep{bluck2020, baker2022}. In \Fig{fig:splashback_kde} we build on this by interpolating between these points using a Kernel Density Estimation (KDE), and include contours to show how the splashback radius varies with cosmology. This shows that the direction in which the splashback radius varies is not constant, but differs across the $\Omega_{\rm{M}} $--$\sigma_8$ plane. 

The values of splashback radius produced by these simulations are generally consistent with others in the literature. \citet{diemer2014} find that high-mass haloes have a splashback feature at $r_{\rm{sp}}\approx2R_{200\rm{c}}$. More recently, \citet{oneil2021} showed that the average splashback radius of clusters in the Illustris TNG300 simulation \citep{nelson2018} is $1.3\pm0.1R_{200\rm{m}}$. Assuming a typical ratio between $R_{200\rm{m}}$ and $R_{200\rm{c}}$ of 1.6 \citep[e.g.][]{deason2020}, this is equivalent to \mbox{$r_{\rm{sp}}=2.1\pm0.1R_{200\rm{c}}$}, consistent with our simulation whose cosmology is most similar to that in TNG300 \citep[see also][for a selection of other relevant studies]{adhikari2021,contigiani2021,towler2024}. Analytic studies of the splashback radius similarly agree with our results; \citet{shi2016} show that slowly- and moderately-accreting haloes in a Universe with \mbox{$\Omega_{\rm{M}}=0.3$} have a splashback radius of approximately $r_{\rm{sp}}=2R_{200\rm{c}}$. They also show that $r_{\rm{sp}}$ decreases slightly with increasing $\Omega_{\rm{M}}$, as we show in \Fig{fig:splashback_scatter} and \Fig{fig:splashback_kde}.

As shown in \Fig{fig:density_profiles}, the combination of \omsig\ in the $S_{8}$ parameter does not correlate with the splashback feature. \Fig{fig:splashback_kde} also shows how the contours of constant $r_{\rm{sp}}$ are almost perpendicular to those of constant $S_{8}$. However, we find that the ratio of $\sigma_{8}$ to $\Omega_{\rm{M}}$ correlates very strongly with the average splashback radius of a cluster. Empirically, we find a linear relationship between these two, such that

\begin{equation}
    \frac{r_{\rm{sp}}}{R_{200}}\approx a+b\left(\frac{\sigma_{8}}{\Omega_{\rm{M}}}\right)\,.
\label{eq:splashback}
\end{equation}

Here, \mbox{$a=1.71\pm0.02$} and \mbox{$b=0.113\pm0.007$}, although these two parameters are strongly correlated (Pearson correlation coefficient $\rho_{a,b}=-0.95$, covariance ${\rm{Cov}}(a,b)=-1.6\times10^{-4}$). These values were found by fitting \Eq{eq:splashback} using standard $\chi^{2}$ minimization, where the uncertainty in each value of $r_{\rm{sp}}$ comes from fitting a quadratic to the splashback feature, as described in \Sec{sec:splashback}. We also fitted a variant of \Eq{eq:splashback} to the data, with the exponents of \omsig\ as two additional free parameters, but this large number of degrees of freedom resulted in very strongly correlated parameters. The best fit returned powers of unity on both \omsig, and so here we fix these values and fit the equation with only two free parameters, for ease of interpretability.

\begin{figure}
\centering
\includegraphics[width=\columnwidth]{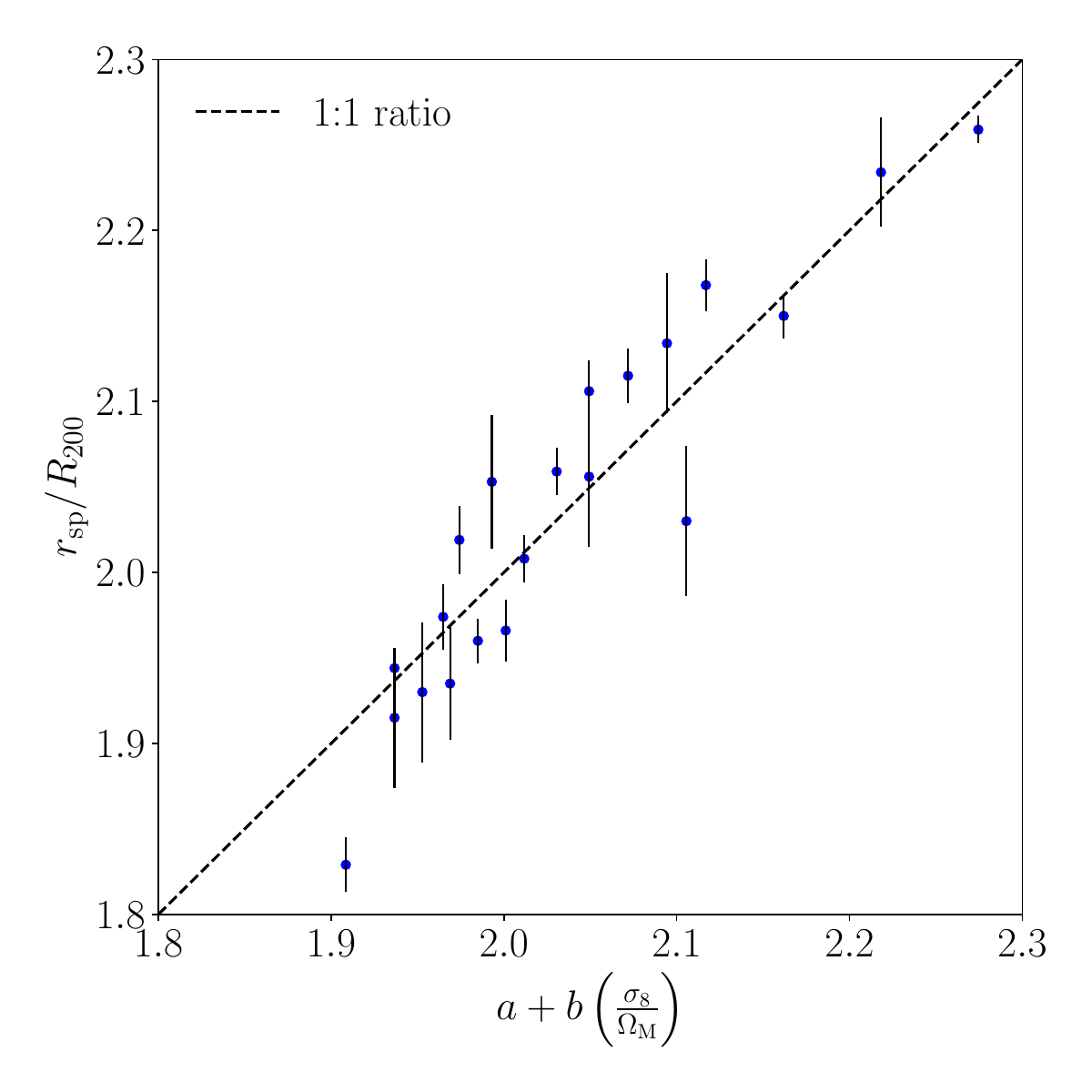}
\caption{Splashback radius, $r_{\rm{sp}}$, in units of $R_{200}$, against model of splashback radius from \Eq{eq:splashback}, where \mbox{$a=1.71\pm0.02$} and \mbox{$b=0.113\pm0.007$}. Black dashed line shows a 1:1 relationship (note that this is slightly different from a line of best fit). Although this relationship is only empirical, it shows how strongly the ratio of $\sigma_{8}$ to $\Omega_{\rm{M}}$ can predict the average splashback radius of clusters.}
\label{fig:r_sp_sig_om}
\end{figure}

The Pearson correlation coefficient between $r_{\rm{sp}}$ and $\sigma_{8}/\Omega_{\rm{M}}$ (which quantifies linear relationships) is equal to 0.93, and the Spearman's rank (which describes monotonic relationships) is also equal to 0.93, indicating a strong, linear, monotonic correlation. This can be seen in \Fig{fig:r_sp_sig_om}. Consequently, a measurement of $r_{\rm{sp}}$ can be used as a proxy for the ratio of \omsig.

\Fig{fig:r_sp_sig_om} shows that there is a large scatter in this relationship. Additionally, we calculate a reduced $\chi^{2}$ for these data equal to 4.4, indicating that the uncertainty in each measurement of $r_{\rm{sp}}$ from the simulations does not fully account for this scatter. To explain this, we emphasize that this fit is purely empirical, and so a perfect correlation between these quantities would not be expected. The function in \Eq{eq:splashback} has been chosen for its simplicity, rather than to perfectly explain the relationship between splashback radius and cosmology. 

\subsection{Observability of splashback feature}
\label{sec:observe}

The approach taken in the previous section is very much a theoretical view of the splashback radius, based on the full 3D information available in the simulations. Observationally, this 3D splashback feature will be projected into two dimensions, potentially obscuring the strong feature we find in our simulations. Furthermore, scaling the splashback radius by \mbox{$R_{200}$} requires accurate, unbiased measurements of individual cluster radii (and hence masses), which is challenging observationally.

To address this, we carry out our analysis with the cluster density profiles projected into two dimensions -- that is, the position of each particle was projected along a single axis (arbitrarily chosen as the $z$-axis within our simulations), to represent the projected density profile that is available observationally. Particles within $20\,h^{-1}\,$Mpc of each cluster's center along the $z$-axis were included in the projected profiles. We also do not scale by $R_{200}$, and instead calculate the average splashback radius of clusters in units of $h^{-1}\,$Mpc, using 400 logarithmically-spaced radial bins between $0.01\,h^{-1}\,$Mpc and $20\,h^{-1}\,$Mpc. The physical size of this projected splashback radius, in units of $h^{-1}\,$Mpc, can be measured observationally using only the angular diameter distance to a cluster and its angular splashback radius (we will discuss practial implementations of observational tests further in Mpetha et al.~in prep).

\begin{figure*}
\includegraphics[width=\textwidth]{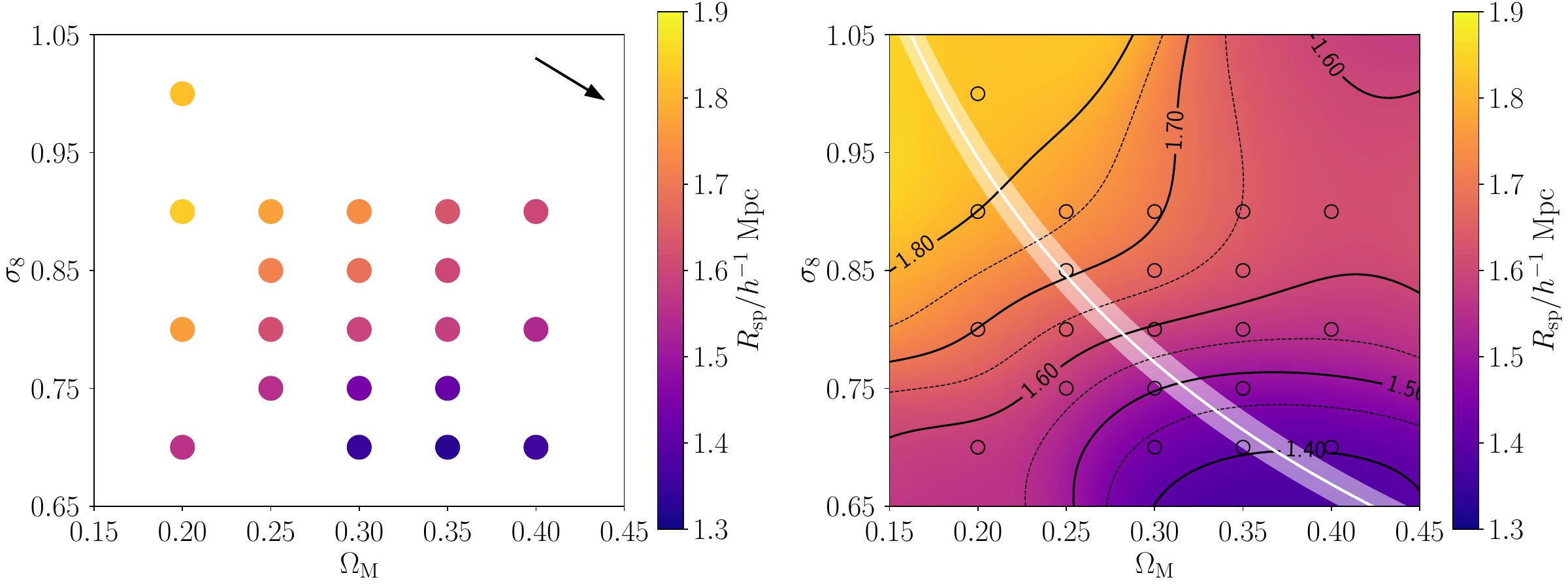}
\caption{Left and right panels are equivalent to \Fig{fig:splashback_scatter} and \Fig{fig:splashback_kde} respectively, but with splashback radii projected into two dimensions, and given in units of $h^{-1}\,$Mpc. The white shaded region in the right panel shows the values of \omsig\ corresponding to a recent measurement of $S_{8}$, of $0.772^{+0.018}_{-0.017}$, calculated from Dark Energy Survey data using cosmic shear \citep{amon2022}.}
\label{fig:splashback_projected}
\end{figure*}

\Fig{fig:splashback_projected} is analogous to the two plots in \Fig{fig:splashback_scatter} and \Fig{fig:splashback_kde}, but with our cluster simulations projected into 2D -- the 2D splashback radius is instead denoted by $R_{\rm{sp}}$, as opposed to $r_{\rm{sp}}$ which was used in 3D. The average density profile is given in units of $h^{-1}\,$Mpc. The direction in which the splashback radius varies across the $\Omega_{\rm{M}}$-$\sigma_{8}$ plane has changed slightly, becoming slightly steeper. This can be interpreted as a slightly stronger dependence on $\sigma_{8}$, and a slightly weaker dependence on $\Omega_{\rm{M}}$ compared to in \Fig{fig:splashback_scatter} and \Fig{fig:splashback_kde}. However, generally the correlation between the splashback radius and these cosmological parameters is maintained in these plots. For comparison, the white band on this plot shows competitive current constraints on $S_{8}$ from cosmic shear measurements, converted to constraints on \omsig\ using \Eq{eq:s8}.

When given in physical units instead of units of $R_{200}$, the splashback radius is no longer well-described by \Eq{eq:splashback}. Instead, a modified version of this empirical formula provides a far better approximation:

\begin{equation}
    \frac{R_{\rm{sp}}}{h^{-1}\rm{Mpc}}=a'+b'\left(\frac{\sigma_{8}}{\Omega_{\rm{M}}^{0.3}}\right)\,.
\label{eq:splashback_projected}
\end{equation}

Similarly to in \Eq{eq:splashback}, the exponents of 1 and 0.3 on $\sigma_{8}$ and $\Omega_{\rm{M}}$ respectively were originally determined by leaving these as free parameters. These were then fixed, to more accurately determine $a'$ and $b'$ in this empirical fit. 

The power of 0.3 on $\Omega_{\rm{M}}$ reflects the statement above, that the splashback radius in units of $h^{-1}\,$Mpc is more strongly dependent on $\sigma_{8}$. The best-fit for the data against this model gives \mbox{$a'=0.5\pm0.1$} and \mbox{$b'=0.95\pm0.09$}, although the parameters $a'$ and $b'$ are again strongly covariant, as in \Eq{eq:splashback} (Pearson correlation coefficient $\rho_{a',b'}=-0.99$, covariance \mbox{${\rm{Cov}}(a',b')=-0.0097$}). 

Finally, the empirical model of \Eq{eq:splashback_projected} provides a good approximation for $R_{\rm{sp}}$ based on \omsig; we find a Pearson correlation coefficient between $R_{\rm{sp}}$ and $\sigma_{8}/\Omega_{\rm{M}}^{0.3}$ equal to 0.93, and a Spearman's rank of 0.98, once again showing the strong, linear relationship between these two quantities.

\begin{figure}
\centering
\includegraphics[width=\columnwidth]{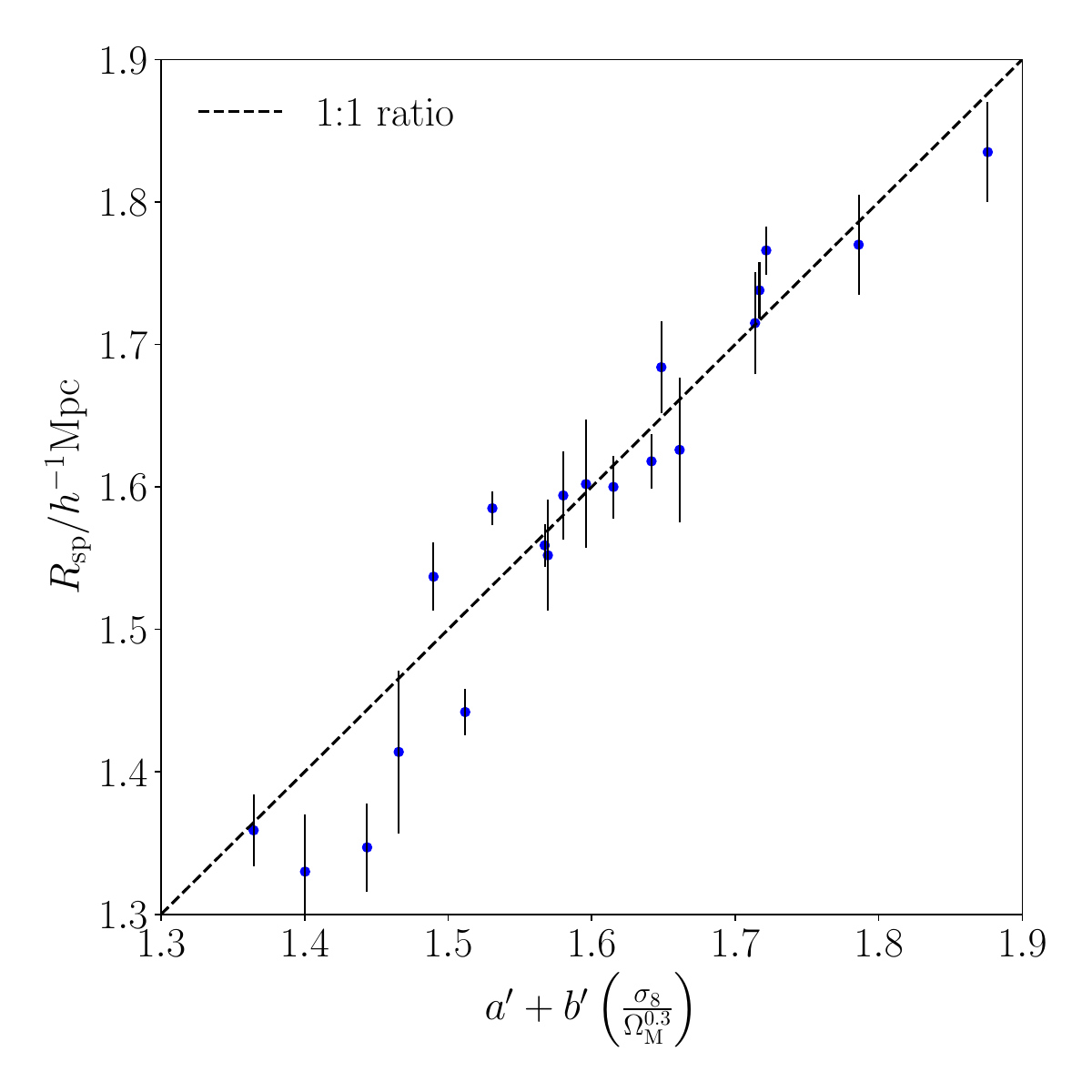}
\caption{Projected splashback radius, $R_{\rm{sp}}$, in units of $h^{-1}\,$Mpc, against model of splashback radius from \Eq{eq:splashback_projected}, where \mbox{$a=0.5\pm0.1$} and \mbox{$b=0.95\pm0.09$}. Black dashed line shows a 1:1 relationship.}
\label{fig:r_sp_projected}
\end{figure}

\subsection{Dependence on halo mass function}
\label{sec:halo_mass_function}

It is well-established that different combinations of \omsig\ will lead to different halo mass functions. These mass functions describe the number density of dark matter haloes as a function of halo mass, and generally decrease sharply at high masses, representing the low number of large haloes in the Universe, compared to the large number of small haloes. Greater values of \omsig\ lead to a greater number of large cluster-sized haloes; it is this connection that allows cluster abundance studies to constrain the value of $S_{8}$ (\Eq{eq:s8}). The ``amplitude'' of the halo mass function will not matter in our study, as we are assuming a complete sample of haloes with masses greater than $10^{14}$, and so the size of such a sample will not impact a measurement of the average splashback radius. However, if the slope of the mass function changes between cosmologies, this could represent a potential issue in our study. 

Previous work has shown that more massive clusters generally have slightly lower splashback radii than smaller clusters, as a fraction of $R_{200}$ \citep{diemer2014,oneil2021}. Consequently, if one combination of \omsig\ produces a flatter halo mass function, a greater fraction of the clusters will be of high mass, which would cause the average splashback radius to decrease.

We study the impact of the different halo mass functions on our results by focusing on one of our simulations in particular, with $\Omega_{\rm{M}}=0.2$ and $\sigma_{8}=1.0$. In \Fig{fig:splashback_scatter} we show that \mbox{$r_{\rm{sp}}=2.26\pm0.01R_{200}$} for haloes of mass greater than $10^{14}\,h^{-1}\,M_{\odot}$ in this simulation. This combination of \omsig\ has the flattest halo mass function of our 21 cosmologies, meaning the most ``top-heavy'' mass function -- we find a (very weak) overall trend, where lower values of $\Omega_{\rm{M}}$ and greater values of $\sigma_{8}$ produce a flatter halo mass function. To test the impact of this flattened halo mass function, we artificially steepened its mass function by stochastically removing haloes from this simulation, with a bias towards more massive haloes, to match the slope of the mass function in another of our simulations. We repeated this 20 times, matching the mass function in this cosmology to each of the 20 other cosmologies in turn. We then re-calculated the average splashback radius of this reduced cluster population in the $\Omega_{\rm{M}}=0.2$, $\sigma_{8}=1.0$ simulation. 

We found that matching the halo mass functions from the other cosmologies had little impact on the average splashback radius calculated in the $\Omega_{\rm{M}}=0.2$, $\sigma_{8}=1.0$ simulation. The 20 newly calculated values ranged from \mbox{$2.25\pm0.01R_{200}$} to \mbox{$2.31\pm0.02R_{200}$} -- this is not a significant difference from the actual value of \mbox{$2.26\pm0.01R_{200}$}, and represents a very small change compared to the variation between cosmologies. Consequently, we conclude that the variation in the halo mass function with \omsig\ does not explain the results we present in this work.

\subsection{Dependence on baryonic physics}
\label{sec:baryonic}

The results presented in this study use a suite of dark matter-only cosmological simulations, based on the assumption that baryonic processes do not have a significant impact on galaxy clusters, due to domination of dark matter over baryonic material on cluster scales. However, it is well-established that baryonic processes can have some impact on the properties of dark matter haloes. This impact is thought to be particularly strong in the centers of clusters, and so not so impactful in the less dense splashback region of a cluster \citep{haggar2021}. 

Nevertheless, as the purpose of the results in this paper is the help constrain properties of the real Universe, we also investigate the difference in these results when baryonic physics is included or excluded. To do this, we use a different group of simulations. \threehun\ project (hereafter \tth) is a suite of 324 zoom simulations of individual galaxy clusters taken from the dark-matter-only MDPL2 MultiDark simulation \citep{klypin2016}. Each of these clusters is simulated from its initial conditions, but with increased resolution and baryonic physics included, giving a mass-complete sample at $z=0$. A far more extensive description of these simulations can be found in \citet{cui2018}.

This suite of 324 clusters have been simulated using multiple different physics codes, and we use the iteration of these simulations run with the {\sc{gadgetX}} code, which is a modified version of the {\sc{gadget-3}} code \citep{springel2005,beck2016}. Specifically, we utilize two versions of these simulations: one including baryonic physics, and one containing only dark matter, both of which use the same initial conditions and the same cosmological parameters \citep[\mbox{$\Omega_{\rm{M}}=0.307$}, \mbox{$\Omega_{\rm{\Lambda}}=0.693$}, \mbox{$h=0.678$}, \mbox{$\sigma_{8}=0.823$};][]{planck2016}. This allows us to study the splashback radius in two samples of galaxy clusters that are identical apart from the inclusion/exclusion of baryonic physics, to investigate the validity of our dark matter-only simulations throughout this work. 

We find that the average splashback radius of clusters is very similar between the dark matter-only simulations and the hydrodynamical simulations. In the dark matter-only simulations, the average splashback radius is equal to \mbox{$1.907\pm0.004R_{200}$}. When baryons are included, this decreases to \mbox{$r_{\rm{sp}}=1.882\pm0.003R_{200}$}. While this difference is significant ($\sim5\sigma$), it only corresponds to a $1\%$ variation in the average splashback radius, compared to the much larger ($>10\%$) variation due to \omsig\ shown above. Consequently, while the predictions of this paper will need minor calibration and correction using hydrodynamical simulations, they are still provide a viable new test in observational cosmology. 

According to \Fig{fig:splashback_kde}, \Fig{fig:r_sp_sig_om} and \Eq{eq:splashback}, the expected splashback radius for clusters in \tth\ simulations would be $2.01\pm0.03R_{200}$, based on the cosmological parameters used in \tth. Similarly, in the dark matter simulation with cosmology most similar to \tth\ ($\Omega_{\rm{M}}=0.3$, $\sigma_{8}=0.8$), the average splashback radius we find is $2.01\pm0.01R_{200}$. This is slightly greater than what is actually found in \tth, which is likely due to the different cluster masses used. \tth\ clusters vary in mass, $M_{200}$, from \mbox{$5\times10^{14}\ h^{-1}M\odot$} to \mbox{$2.6\times10^{15}\ h^{-1}M\odot$}. Meanwhile, the dark matter simulations we use throughout this study include all clusters with masses down to $10^{14}\ h^{-1}M\odot$, and as we described in the previous section, lower-mass haloes are knwon to have greater splashback radii. Indeed, if we repeat our analysis using only clusters with \mbox{$M_{200}>5\times10^{14}\ h^{-1}M\odot$}, we find that the average splashback radius in our $\Omega_{\rm{M}}=0.3$, $\sigma_{8}=0.8$ simulation decreases to $1.83\pm0.09R_{200}$, consistent with the average splashback radius in the dark matter-only simulation from \tth\ suite.

\section{Discussion}
\label{sec:discussion}

This work shows that the mean splashback radius of large galaxy clusters can be used as a proxy for measuring cosmological parameters, namely \omsig. The variation of $r_{\rm{sp}}$ with these parameters is almost perpendicular to the variation of $S_{8}$, the parameter best constrained by low-redshift measurements of structure formation -- that is, many values of $r_{\rm{sp}}$ correspond to a single value of $S_{8}$. Similarly, measurements of cluster abundance or cosmic shear result in a strong degeneracy between \omsig\ \citep{abdullah2020,amon2022}.

This can be clearly seen in \Fig{fig:splashback_projected}. The white line in the right panel of this plot shows the values of \omsig\ corresponding to a recent constraint on $S_{8}$ (with $1\sigma$ uncertainty), using cosmic shear measurements from the Dark Energy Survey \citep{amon2022}. This contour runs almost perpendicular to the lines of constant splashback radius. Consequently, combining a precise measurement of the splashback radius with a measurement of $S_{8}$ would break the degeneracy between \omsig, allowing precise values for both of these properties to be measured at once.

The correlation between $r_{\rm{sp}}/R_{200}$ and these cosmological parameters can be interpreted similarly to the relationship between splashback radius and the formation time of galaxy clusters. Previous studies have established that the splashback feature around recently-formed, dynamically disturbed clusters is close to $R_{200}$, and in dynamically relaxed clusters it lies further from the cluster center. \citet{shin2023} show that clusters with earlier formation times, and clusters with lower present-day accretion rates, both have greater splashback radii at $z=0$. Similarly, \citet{kuchner2022} study the backsplash population of galaxies around simulated clusters (analogous to the splashback radius), and show that this population are found at greater distances around dynamically relaxed clusters. In a related study, \citet{haggar2020} show a greater fraction of backsplash galaxies surrounding clusters that formed long ago, compared to recently-formed clusters.

The reason for this is that the splashback feature is a probe of the growth history of a galaxy cluster. The splashback feature is made up of material that was accreted by the cluster approximately half an orbital period in the past, and has now reached the apocenter of its orbit \citep{diemer2014}. The potential of the cluster several gigayears ago (at the epoch of infall) drives the infall speed of material, which in turn determines the orbit of the material and thus its apocentric distance. In clusters that have formed recently (acquired much of their mass at recent times), the potential in the past was smaller, meaning the infall speed was low, and so the splashback radius at the present day is not far out from the cluster center \citep[see also][]{shin2023}.

In universes with differing values of \omsig, the average time at which large-scale structure collapses and clusters form also varies. \citet{amoura2021} use a similar suite of simulations as in this study, but with only nine combinations of \omsig, and a lower mass resolution. They show that for cosmologies in the upper-left corner of the $\Omega_{\rm{M}}$-$\sigma_{8}$ plane (low $\Omega_{\rm{M}}$, high $\sigma_{8}$), galaxy clusters typically form at earlier times (a median formation time, $z_{0.5}$ of $0.8$, compared to a formation time of $0.5$ for cosmologies with high $\Omega_{\rm{M}}$ and low $\sigma_{8}$). These earlier-forming clusters are equivalent to the dynamically relaxed clusters discussed in other works, and indeed we find that clusters in the upper-left of these plots do have a greater ratio of $r_{\rm{sp}}$ to $R_{200}$. Studies on smaller scales have found similar results; \citet{conselice2014} show that models with low $\Omega_{\rm{M}}$ and high $\sigma_{8}$ produce a lower merger rate of galaxy-mass dark matter haloes, indicating that the haloes will be more dynamically relaxed on average. We will investigate this further in an upcoming paper (Amoura et al.~in prep), by studying the variation in accretion history and present-day accretion rate with cosmology. This will build on the previous work of \citet{amoura2024}, who examined the instantaneous growth rate of clusters in a subset of the simulations used in our work; their numerical and analytical results both support the idea that cosmologies in the bottom-right of the $\Omega_{\rm{M}}-\sigma_{8}$ plane lead to clusters growing more rapidly at recent times. 

Moreover, we can explain this relationship by considering the parameters \omsig\ individually. Cosmologies with a low value of $\Omega_{\rm{M}}$ will have a greater value of the cosmological constant, $\Omega_{\Lambda}$, as all of these simulations are of a flat universe. The present-day Hubble parameter in these simulations is also fixed, at a value of $70\,$km$\,$s$^{-1}\,\rm{Mpc}$, meaning that the Hubble parameter at higher redshifts, $H(z)$, will be less in simulations with lower $\Omega_{\rm{M}}$ and greater $\Omega_{\Lambda}$. Consequently, in these simulations, the turnaround radius of a cluster (the distance where the Hubble flow outmatches the gravitational pull of a cluster) will be pushed further out. This will lead to more material falling into the cluster, and thus the gravitational collapse of structure occurring at earlier times (as shown in \citealt{amoura2021}). Additionally, a greater value of $\Omega_{\Lambda}$ means that the matter-dominated phase of the Universe -- when much of the large-scale structure collapses -- will end sooner, also resulting in clusters forming at earlier times. Clusters in these low-$\Omega_{\rm{M}}$, high-$\Omega_{\Lambda}$ simulations will therefore have more time to become dynamically relaxed, explaining their greater splashback radius.

Similarly, a greater value of $\sigma_{8}$ also impacts the formation of structure. An increased amplitude of matter density fluctuation on the scale of clusters accelerates the collapse of these structures, and so means that the average formation time of galaxy clusters is earlier in these simulations. A greater value of $\sigma_{8}$ will therefore have a similar effect to a low value of $\Omega_{\rm{M}}$, providing clusters with more time to become dynamically relaxed, and so a greater splashback radius, as we see in \Fig{fig:splashback_scatter}.

\section{Conclusions}
\label{sec:conclusions}

In this work, we use a suite of dark matter-only cosmological simulations to show that the location of the splashback radius around large galaxy clusters depends on the ratio of \omsig. Specifically, in a cosmology where $\Omega_{\rm{M}}$ is lower and $\sigma_{8}$ is greater, the splashback radius will be pushed outwards. This effect is present in both 3D simulation data, and in a 2D projection, and is not strongly dependent on baryonic physics in the clusters. In \Sec{sec:discussion}, we discuss how this increased splashback radius is connected to a greater average apocentric distance of bound cluster material, and also to the dependence of cluster formation times and dynamical states on cosmological parameters.

This work consequently demonstrates a novel way to estimate the cosmological parameters \omsig, using a measurable property, the splashback radius. The splashback radius can be measured from weak lensing data \citep[e.g.][]{chang2018}, which is already widely used to constrain $S_{8}$ in cosmic shear measurements. Through the measurement of the splashback radius, the existing degeneracy between \omsig\ can be broken without the need for new observational data, constraining low-redshift estimates of these cosmological parameters to a smaller region of the $\Omega_{\rm{M}}$-$\sigma_8$ plane.

To apply this study to observations, precise measurements of the average splashback radius of complete samples of massive \mbox{($>10^{14}\,M_{\odot}$)} clusters are needed -- to make competitive constraints, a precision of less than $5\%$ in the mean will be required. Furthermore, the mass dependence we discuss in \Sec{sec:halo_mass_function} and \Sec{sec:baryonic} demonstrates the necessity of matching an observational cluster sample to those used in studies like ours, as biased cluster mass measurements will lead to a biased measurement of the splashback radius. In a follow-up study (Mpetha et al.~in prep), we will examine the feasibility of constraining \omsig using measurements of the splashback radius and other cluster features in the infall region outside the virial radius, focusing on specific surveys such as {\textit{Euclid}}\footnote{\url{https://www.esa.int/Science_Exploration/Space_Science/Euclid}} and UNIONS\footnote{\url{https://www.skysurvey.cc/}}. 

The splashback radius of smaller (group or galaxy-mass haloes) is less pronounced, which is why we focus on galaxy clusters in this study. Smaller haloes, however, have a more pronounced depletion zone in their outer regions \citep[see][]{Fong2021}. If measurable, this would complement measurements of the splashback feature in clusters; again, we will discuss using the depletion zone as a cosmological test in a follow-up study (Mpetha et al.~in prep).

\begin{acknowledgments}

CTM is funded by a Leverhulme Trust Study Abroad Scholarship. JET acknowledges financial support from NSERC Canada, through a Discovery Grant. This research was enabled in part by support provided by Compute Ontario (www.computeontario.ca) and the Digital Research Alliance of Canada (alliancecan.ca).

This work makes use of the {\sc{SciPy}} \citep{virtanen2020}, {\sc{NumPy}} \citep{vanderwalt2011}, {\sc{Matplotlib}} \citep{hunter2007} and {\sc{pandas}} \citep{mckinney2010} packages for {\sc{Python}}.

\end{acknowledgments}

%

\vspace{5mm}





\appendix

\section{Splashback radius fitting}
\label{appendix:splashback_fitting}

For each of the 21 dark matter-only simulations used in this work, the average splashback radius for cluster-sized haloes was calculated from the average slope of the density profiles of these haloes. The average density profile and the slope of this profile for each of the 21 combinations of \omsig\ are shown in the left and right panels of \Fig{fig:density_profiles} respectively, smoothed with a Savitsky-Golay filter \citep{savitsky1964}. 

\begin{figure*}
\includegraphics[width=\textwidth]{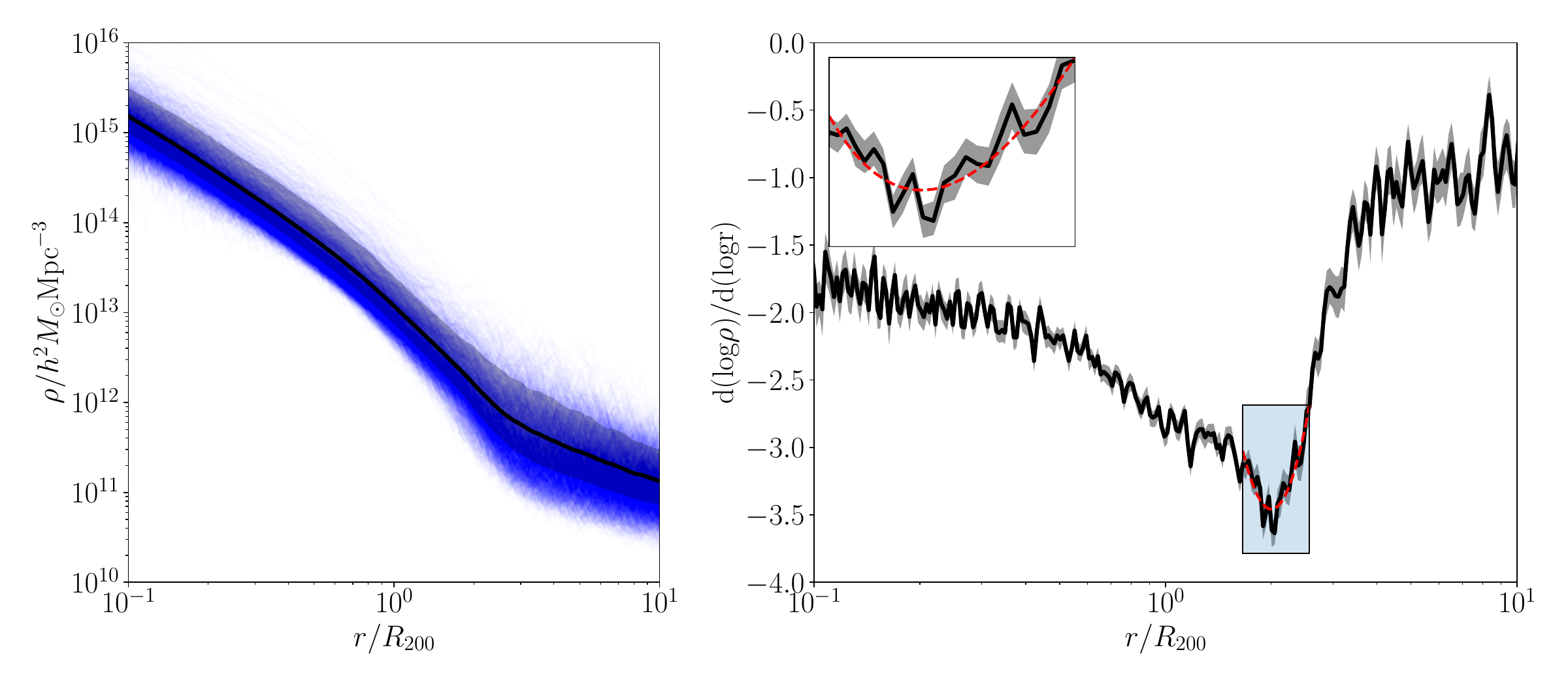}
\caption{Left panel shows the radial density profile for all cluster-sized haloes ($M_{200}>10^{14}\,h^{-1}\,M_{\odot}$) in the dark matter-only simulation with $\Omega_{\rm{M}}=0.3$ and $\sigma_{8}=0.8$, using 3D (not projected) data. These individual profiles are given by the thin blue lines, and the thick black line shows the mean of these density profiles as a function of radius. Dark shaded region shows the $1\sigma$ spread of these data; the uncertainty in the mean, calculating by bootstrapping, is too small to be visible on this scale. The right panel shows the slope of this mean density profile as a function of radius, and the shaded region shows the uncertainty in this. The zoomed, highlighted region around the splashback feature shows the curve that was fitted to this minimum (red dashed line) to calculate the splashback radius. In this case, that splashback radius is equal to \mbox{$r_{\rm{sp}}=2.00\pm0.01R_{200}$}. For clarity, the slopes of the individual haloes are not shown.}
\label{fig:m3s8_appendix}
\end{figure*}

The unsmoothed versions of these plots were used to calculate the average cluster splashback radius independently in each of the 21 cosmologies. \Fig{fig:m3s8_appendix} shows a typical example of this, with the left panel showing the radial density profiles of all cluster-sized haloes in the dark matter-only simulation with $\Omega_{\rm{M}}=0.3$ and $\sigma_{8}=0.8$. The average (mean) of these, and the $1\sigma$ spread in these data are also shown. While the complete density profiles consist of 400 radial bins between $0.01\,R_{200}$ and $20\,R_{200}$, as detailed in \Sec{sec:splashback}, here we only show data between $0.1\,R_{200}$ and $10\,R_{200}$, equivalent to approximately 242 radial bins.

The right panel of this plot shows the slope of the mean density profile in log-log space, as in \Fig{fig:density_profiles}, and an uncertainty in this average calculating by bootstrapping the data. In the region surrounding the splashback feature, we fit a quadratic curve to the data (red dashed line), and use the minimum of this as the value of the splashback radius. A zoomed view of this region is inset to the right panel of this plot. In this case, we calculate the location of this minimum to be \mbox{$r_{\rm{sp}}=2.00\pm0.01R_{200}$}, with a reduced $\chi^{2}$ statistic consistent with the number of degrees of freedom ($\chi_{\rm{\nu}}^{2}=1.03$).

\bibliography{main}{}
\bibliographystyle{aasjournal}



\end{document}